\def\jcap{JCAP}
\def\mnras{MNRAS}             
\def\aap{A\&A}             

\documentclass[%
reprint,
superscriptaddress,
nofootinbib,
nolongbibliography,
 amsmath,amssymb,
 aps,
prd,
floatfix,
]{revtex4-2}

\usepackage{graphicx}
\usepackage{dcolumn}
\usepackage{bm}
\usepackage{xcolor}

\definecolor{maroon}{RGB}{192,0,0}
\usepackage[colorlinks=true, citecolor=maroon,
linkcolor=maroon,
urlcolor=blue]{hyperref}
\usepackage{times}
\usepackage{tikz}

\def\checkmark{\tikz\fill[scale=0.4](0,.35) -- (.25,0) -- (1,.7) -- (.25,.15) -- cycle;} 

\newcommand{\karmma}{{\sc karmma}~}
\newcommand{\healpix}{\textsc{Healpix}}
\newcommand{\Nside}{N_{\rm side}}
\newcommand{\gansky}{{\sc gansky}~}

\begin{document}

\title{GANSky -- fast curved sky weak lensing simulations using Generative Adversarial Networks}

\author{Supranta S. Boruah}
 \email{supranta@sas.upenn.edu}
\affiliation{Department of Astronomy and Steward Observatory, University of Arizona, 933 N Cherry Ave, Tucson, AZ 85719, USA}
\affiliation{Department of Physics and Astronomy, University of Pennsylvania, Philadelphia, PA 19104, USA}
\author{Pier Fiedorowicz}
\affiliation{Department of Physics, University of Arizona, 1118 E. Fourth Street, Tucson, AZ, 85721, USA}
\affiliation{Lawrence Livermore National Laboratory, Livermore, CA 94550, USA}
\author{Rafael Garcia}
\affiliation{Department of Physics, University of Arizona, 1118 E. Fourth Street, Tucson, AZ, 85721, USA}
\author{William R. Coulton}
\affiliation{Kavli Institute for Cosmology Cambridge, Madingley Road, Cambridge CB3 0HA, UK}
\affiliation{DAMTP, Centre for Mathematical Sciences, University of Cambridge, Wilberforce Road, Cambridge CB3 OWA, UK}
\author{Eduardo Rozo}
\affiliation{Department of Physics, University of Arizona, 1118 E. Fourth Street, Tucson, AZ, 85721, USA}
\author{Giulio Fabbian}
\affiliation{Kavli Institute for Cosmology Cambridge, Madingley Road, Cambridge CB3 0HA, UK}
\affiliation{Institute of Astronomy, Madingley Road, Cambridge CB3 0HA, UK}
\affiliation{School of Physics and Astronomy, Cardiff University, The Parade, Cardiff, CF24 3AA, UK}
\affiliation{Center for Computational Astrophysics, Flatiron Institute, New York, New York 10010, USA}

\date{\today}

\begin{abstract}
Extracting non-Gaussian information from the next generation weak lensing surveys will require fast and accurate full-sky simulations. This is difficult to achieve in practice with existing simulation methods: ray-traced $N$-body simulations are computationally expensive, and approximate simulation methods (such as lognormal mocks)  are not accurate enough. Here, we present {\sc gansky}, an interpretable machine learning method that uses Generative Adversarial Networks (GANs) to produce fast and accurate full-sky tomographic weak lensing maps. The input to our GAN are lognormal maps that approximately describe the late-time convergence field of the Universe. Starting from these lognormal maps, we use GANs to learn how to locally redistribute mass to achieve simulation-quality maps.  This can be achieved using remarkably small networks ($\approx 10^3$ parameters). We validate the GAN maps by computing a number of summary statistics in both simulated and \gansky\ maps. We show that \gansky\ maps correctly reproduce both the mean and $\chi^2$ distribution for several statistics, specifically: the 2-pt function, 1-pt PDF, peak and void counts, and the equilateral, folded and squeezed bispectra.  These successes makes \gansky an attractive tool to compute the covariances of these statistics. In addition to being useful for rapidly generating large ensembles of artificial data sets, our method can be used to extract non-Gaussian information from weak lensing data with field-level or simulation-based inference.
\end{abstract}

\maketitle


\section{Introduction}\label{sec:intro}

Weak lensing is one of the main cosmological probes with which next generation cosmological surveys --- e.g. Vera Rubin Observatory's Legacy Survey of Space and Time \citep[LSST,][]{LSST19}, the Roman Space Telescope \citep{Roman15} and Euclid \citep{euclid11} --- will seek to understand the nature of dark energy and dark matter. Accessing the non-Gaussian information in these weak lensing data sets is expected to  improve the cosmological constraints derived from said data by factors of 2 or more \citep{Boruah2023, Fluri2021, Euclid2023_NG}.   This extraction can be accomplished in a variety of ways.  In field-based inference \citep{Millea2021, Porth2023, Boruah2023, Porqueres2022} and simulation-based inference studies \citep{Taylor2019, Jeffrey2021, Gatti2023, Gatti2024} one compares observations to synthetic data sets to extract cosmological parameter information.  Alternatively, the non-Gaussian information can be accessed using non-Gaussian summary statistics \cite{Euclid2023_NG}, e.g. peak and void counts \citep{Liu2015, Coulton2020, Marques2023}, scattering transforms \citep{Cheng2020, Ajani2021}, and the 1-point PDF \citep{Boyle2021, Thiele2023}. Because these probes are sensitive to non-linear structure formation, the calibration of both the mean and covariance matrix of these summary statistics has to be achieved with synthetic data sets.  That is to say, regardless of the analysis method, fast and accurate simulations are essential for exploiting the non-Gaussian information with weak lensing data.

A number of different simulation methods have been proposed in the literature. As a general rule of thumb, there is a trade-off between accuracy and speed. $N$-body simulations are accurate, but expensive, while approximate methods are fast, but inaccurate. One of the widely used methods for producing approximate simulations of weak lensing data assumes that the convergence field can be described as a lognormal random field \citep{Xavier2016,Fiedorowicz2022a, Fiedorowicz2023, Boruah2022, Boruah2023, Boruah2024}. However, the lognormal model fails at small scales and low redshifts. Other approximate methods based on analytic prescriptions include modeling the density field in redshift shells \citep{Tessore2023} or using an inverse-Gaussianization method \citep{Yu2016}.

Generative machine learning methods present an attractive approach for learning the probability distribution of high-fidelity simulations using neural networks. Once trained, these neural networks can quickly generate high-quality simulations. Indeed, various generative machine learning models have been used for producing fast and accurate simulations in cosmology. Examples of machine learning tools used include Generative Adversarial networks (GAN) \cite{Rodriguez2018, Mustafa2019,Shirasaki2023}, styled neural networks \citep{Jamieson2023}, normalizing flows \citep{Dai2022, Dai2023}, and neural score matching \citep{Remy2023}.

In this paper, we use Generative Adversarial Networks (GAN) to produce full-sky weak lensing simulations. GANs have been successful in modeling complex data sets \citep{biggan}. A GAN is comprised of two neural networks which compete with each other through an adversarial loss. The first network is the generator, which takes a random sample from a latent space prior as input and generates a synthetic data sample. The second network is the discriminator, which learns to distinguish real samples from those created by the generator. The networks are adversarial in that the discriminator is trained to differentiate between real and synthetic samples, while the generator is trained to fool the discriminator. GANs are especially well suited for modeling systems in which real samples are easy to collect but cannot be analytically described. As such, they are ideally suited for modeling the outputs of $N$-body simulations \citep[e.g.][]{Rodriguez2018,KodiRamanah2020, Li2021}. 

We are not the first to propose the use of GANs for generating synthetic weak lensing maps. However, prior work has been limited to either small flat sky patches \cite{Mustafa2019, Shirasaki2023}, or curved-sky patches with relatively small fixed footprints \cite{Yiu2022}. By contrast, current and future wide-field cosmological surveys necessitate generating maps over large fractions of the celestial sphere.

Here, we train our neural networks on the curved sky and use the GAN generator to produce full-sky weak lensing simulations. Our method differs from previous works in several key elements.  First, our generator produces full-sky simulations.  Second, unlike traditional machine learning approaches \citep{Mustafa2019,Yiu2022}, the input to our generator is not an un-interpretable random noise vector, but rather a log-normal realization of the convergence field.  Our generator is then trained so as to deform our input lognormal map into a simulation-quality map using the adversarial process. Similar ideas have already been executed in \cite{Piras2023} to produce projected density fields from $N$-body simulations, and by \cite{Shirasaki2023} within the context of generating weak lensing maps from Gaussian maps.  
However, both of these works treat the process of transforming the log-normal map into a simulation quality map as a general image-deformation processes that ignores the detailed nature of the image at hand.  Consequently, these techniques ignore the strong physical constraint that the images we are interested in represent physical matter density fields governed by gravity.  By contrast, our networks are physics-informed by demanding that our generator be restricted to performing a small local mass redistribution of the lognormal simulation in a rotationally equivariant way (see Section~\ref{sec:gan_design} for details).
The corresponding constraints successfully stabilize the training of our networks while reducing the number of free parameter in the generator by three orders of magnitude relative to previous work. Moreover, we train our networks to simultaneously operate on tomographic maps, enabling us to generate sets correlated tomographic convergence maps.

Compared to previous work, the maps we generate are lower in resolution.  Specifically, throughout this work, we aim to produce simulation-quality \healpix\ maps at $\Nside=512$.  This corresponds to a pixel size of $\sim~7$ arcmin. Below this scale, baryonic effects become significant.  Consequently, even if we were to generate high resolution maps, these maps would fail to adequately describe the real Universe. For this reason, we chose to prioritize our ability to generate full-sky tomographic maps at this comparatively coarse resolution.

This paper is structured as follows: in section \ref{sec:sims}, we describe the simulations and training data used in this work. Section \ref{sec:gan_design} describes the GAN used in this work.  We present our results in section \ref{sec:results} before concluding in section \ref{sec:conclusion}.

\section{Simulations and training data}\label{sec:sims}

We use the publicly available weak lensing simulations of \cite[][hereafter T17]{Takahashi2017} to produce the training data for our GAN. T17 produced a suite of $108$ mock weak lensing catalogues by ray-tracing $N$-body simulations run with cosmological parameters $\Omega_m = 0.279$, $\Omega_b = 0.046$, $h = 0.7$, $\sigma_8 = 0.82$ and $n_s = 0.97$. Each mock simulation consists of full-sky convergence maps in redshift shells separated by $150~h^{-1}$ Mpc.

\begin{figure}
    \centering
    \includegraphics[width=\linewidth]{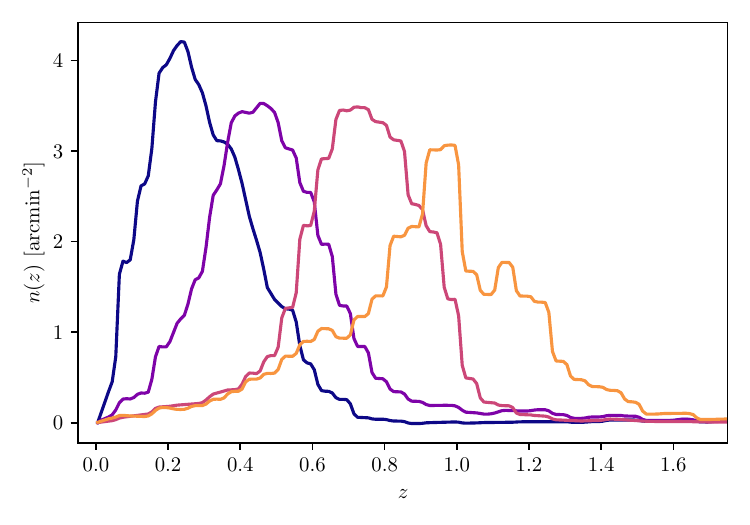}
    \caption{The source redshift distribution, $n(z)$, for the $4$ redshift bins used in this work.}
    \label{fig:nz_plot}
\end{figure}

For a given redshift distribution, $n(z)$, we produce tomographic convergence maps by taking a weighted average of the maps at various redshift shells via
\begin{equation}
    \kappa^{i} = \frac{\sum_{j=1}^{N_{\text{shells}}} n^i(z_j) \kappa_{\text{shell}}^{j}}{\sum_{j=1}^{N_{\text{shells}}} n^i(z_j)},
\end{equation}
where $n^{i}$ is the redshift distribution of the $i$-th tomographic bin and $\kappa^j_{\text{shell}}$ is the convergence map in the $j$-th redshift shell. Note that the finite redshift resolution leads to a systematic error of $\mathcal{O}(5\%)$ on the power spectrum of the convergence field. While we will need to account for this error in future work, here we are simply interested in the performance of the GAN, that is, can our generator successfully return maps that are indistinguishable from the simulated maps. We use the convergence map at a {\sc healpix} resolution of $N_{\text{side}}=512$ to produce our training data. We produce our mock simulations to mimic the galaxy distribution of DES-Y3 as reported by \cite{Myles2021} and shown here in Figure~\ref{fig:nz_plot}.

To train our fiducial GAN, we use a set of 80 full-sky weak lensing simulations. Because our network is restricted to performing a \it local \rm mass redistribution operation, we can train the network using small simulation patches. Consequently, each simulation is further split into 3072 $3.5\times 3.5\ {\rm deg}^2$ regions.  The remaining 28 full-sky simulations are saved for validation. 

\section{GAN design and architecture}\label{sec:gan_design}

\begin{figure*}
    \centering
    \includegraphics[width=\linewidth]{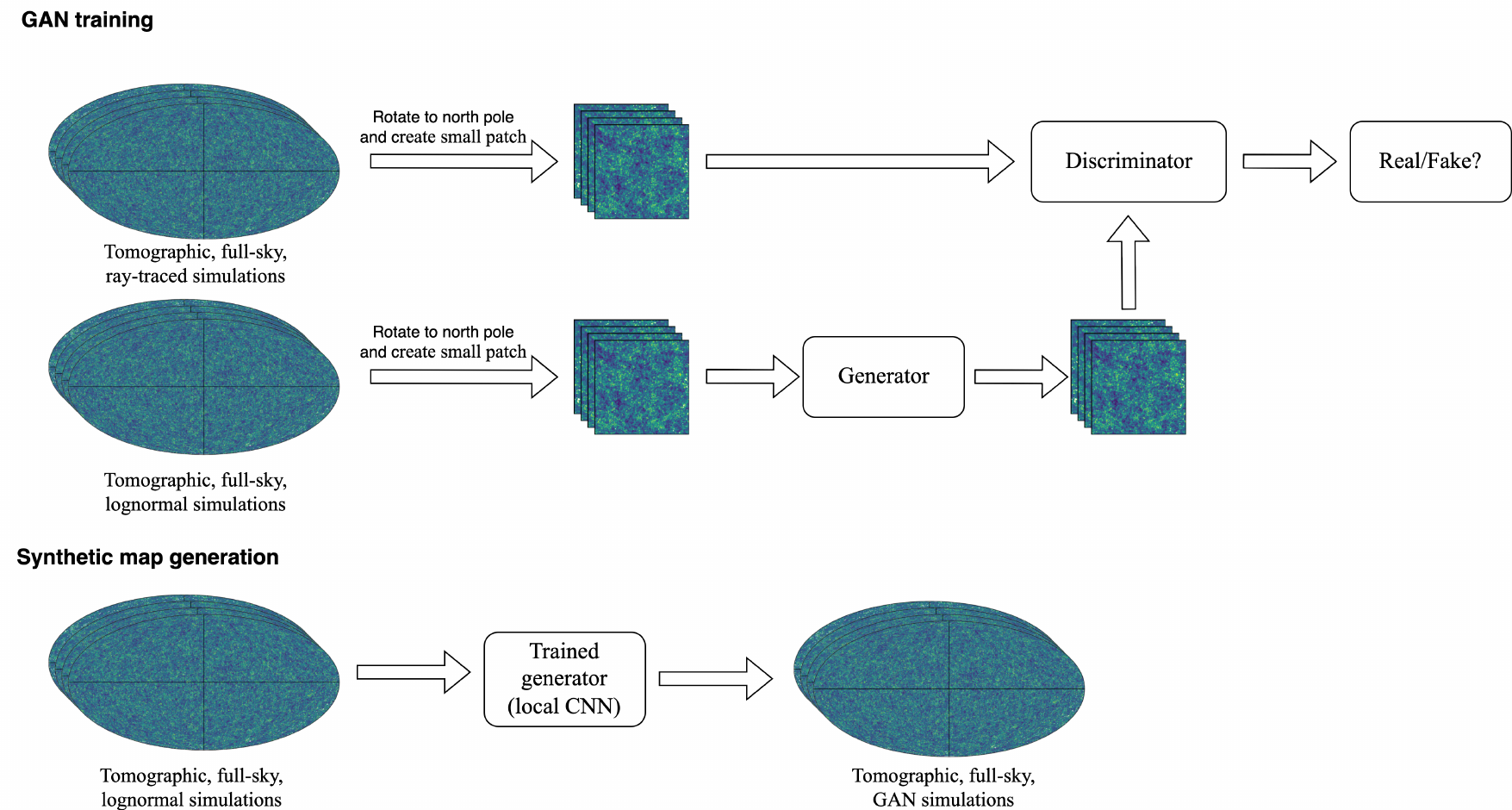}
    \caption{Illustration of our method. For training our GAN, we create small patches from the ray-traced simulation and the lognormal maps. Therefore, our generator only learns the local redistribution of mass in the lognormal maps. The discriminator similarly learns to distinguish the features on these small patches. Once trained, the generator can be used to produce full-sky synthetic maps from the full-sky lognormal simulations. See section \ref{sec:gan_design} for more details.}
    \label{fig:gansky_illustration}
\end{figure*}

\subsection{Our Model at a Glance}\label{ssec:model_glance}

Our method is based on two premises: 1) it is easier to learn the perturbations to simple analytic models than to learn how to generate maps from scratch; and 2) we should be able to deform an approximate analytic map into a simulation quality map via local mass redistribution.  Consequently, we use our GANs to learn how to deform lognormal maps (section~\ref{ssec:lognormal_model}) into simulation-quality maps. We impose three
key properties on our network.  

\begin{enumerate}
    \item \bf Rotational Symmetry: \rm the generator and discriminator rely on rotationally equivariant convolutional neural networks, thereby respecting the rotational symmetry of gravity. We note that rotation-equivariant networks have been shown to perform comparably to or better than anisotropic CNNs \cite{kids_dl, Dai2022, Dai2023} \citep[see also][]{Perraudin2019, Krachmalnicoff2019}. 
    \item \bf Causality: \rm Causality is enforced by demanding that the output value of the convergence field at any given pixel depends only on the pixels in the input map that fall within $\approx 0.5\ {\rm deg}$ of the output pixel.  This radius is roughly a pixel's causal horizon, and sets the number of convolutional layers in our network. 
    \item \bf Network is perturbative: \rm We would like our network to find the simulation-quality map that is ``closest'' to the input lognormal map.  To that end, we add a term to the cost function that penalizes the generator based on how different the input and output maps are. 
\end{enumerate}

The fact that our networks are local implies they can be trained using small ($3.5\times 3.5\ {\rm deg}^2$) patches.  By construction, the resulting generator is a convolutional neural network that operates on \healpix\ maps.  That is, even though it is trained using small patches, the generator can be applied to full-sky lognormal maps with no stitching. The generation of full-sky synthetic maps using the GAN is discussed in section \ref{ssec:synthetic_map_gen}.

Our final network contains $980$ parameters ($936$ parameters for the discriminator).  It takes as input 4 tomographic lognormal maps, and returns four tomographic simulation-quality maps. An illustration of our method is shown in Figure \ref{fig:gansky_illustration}. A comparison of our method with other methods is presented in Table \ref{tbl:model_comparison}. Below, we describe each of the components of our network in further detail.

\begin{table}
  \centering
  \caption{Comparison of the main features of our method with other GAN-based weak lensing map generation methods in the literature. Note our GAN uses orders of magnitude fewer parameters than other works. }
  \begin{tabular}{l | c c c c}
  \hline
      &  This work & \cite{Yiu2022} & \cite{Shirasaki2023} & \cite{Mustafa2019}\\
      \hline
     Input size & $1.2\times10^7$ & $584$ & $256^2$ & $64$\\
     Output size & $1.2\times10^7$ & $3.7 \times 10^{5}$ & $256^2$ & $256^2$\\
     Parameter size & $980$ & $5 \times 10^6$ & $11.4\times10^6$ & $12.3\times10^6$\\
     Curved sky & \checkmark & \checkmark & $\times$ & $\times$\\
     Tomographic maps & \checkmark  & \checkmark & $\times$ & $\times$ \\
    \hline
  \end{tabular}
  \label{tbl:model_comparison}
\end{table}

\subsection{The Lognormal Model}\label{ssec:lognormal_model}

The convergence field from numerical simulations can be approximately described as lognormal \citep{Taruya2002, Clerkin2017}. In this model, the convergence field $\kappa_i$ of the $i$-th tomographic bin is taken to be a non-linear transformation of a Gaussian random field $y_i$. Specifically,
\begin{equation}
    \kappa_i = e^{y_i} - \lambda_i.
    \label{eq:ln}
\end{equation}
The parameter $\lambda_i$ is known as the shift parameter, and is related to the skewness of the probability distribution function (PDF) of $\kappa_i$. 
We fit the shift parameters from ray-traced {\sc healpix} maps.

Multivariate lognormal distributions can be used to model correlated convergence fields in tomographic bins \citep{Xavier2016}. For this model, the correlation function of $y$ between the $i$-th and $j$-th tomographic bins is related to the corresponding $\kappa$ correlation function via 
\begin{equation}
    \xi^{ij}_{yy} = \ln \left( \frac{\xi^{ij}_{\kappa\kappa}}{\lambda_i \lambda_j}+1 \right).
\end{equation}
Given the non-linear power spectrum of the convergence field, we can readily compute the corresponding correlation function of $y$ using the above equation. We then apply a spherical harmonic transform to arrive at the power spectrum of the $y$-field.

We generate our lognormal maps at a resolution of $N_{\text{side}}=2048$. We evaluate the power spectrum $C_l^{yy}$ of the $y$-field at this resolution, up to a maximum $\ell$ of $\ell_{\rm max} = 6144$, and apply the non-linear transformation in equation~\ref{eq:ln} to arrive at a lognormal random field. The power spectrum of the $y$ field is chosen such that the resulting lognormal maps have the same power spectrum as the ray-traced simulations. The high resolution maps are then rotated to the north pole and downgraded to a resolution of $N_{\text{side}} = 512$ to produce the input patch. The details of this process is discussed in section \ref{ssec:training}. For further details on the lognormal model, we refer the reader to \cite{Xavier2016, Boruah2022}.

\subsection{Generative Adversarial Network Implementation}\label{ssec:gan_implementation}

We use GANs to learn how to modify lognormal maps so that they become statistically indistinguishable from those produced from numerical simulations. In a traditional GAN, the dimension of the latent space is much lower than the dimension of the maps.  Moreover, the variables in this latent space are non-interpretable noise variables that are typically assumed to follow a unit Gaussian distribution $z \sim N(\vec{0}, I)$. By contrast, in our method, the latent space has the same dimensionality as the output map from the generator, thereby preserving all of the degrees of freedom in the problem. The latent space prior is that the input map is a random draw from the lognormal model. Because the lognormal model accurately describes the convergence field on large scales,  the action of the generator $G$ on the input map can be thought of as a local mass redistribution, implying convolutional neural networks (CNN) are particularly well suited to our problem. The end result is that despite the high dimensionality of our latent space (our maps contain$\approx 3{\rm M}$~pixels in each tomographic bin), the demands on the generator are drastically reduced relative to the standard approach. Consequently, our network achieves our goals with three order of magnitude fewer parameters than other machine learning approaches.

Based on this discussion, we choose a CNN as our generator. Standard CNNs can only operate on flat Euclidean spaces. To apply convolutions on spherical maps, one option is to split the sphere into small patches, which can be approximated as flat, apply a CNN, and then recombine the patches \citep{cmb_gan}. Alternatively, we can implement CNNs directly on the surface of a sphere using the {\sc healpix} \citep{healpix} pixelization. Here, we use this latter approach and perform rotation equivariant convolutions using the four nearest neighbors to each pixel. Note that this convolution extends over \it all \rm tomographic bins, i.e. a pixel in tomographic bin $i$ has connections to: 1) the same pixel in other tomographic bins $j\neq i$; and 2) the four nearest neighbors in \it all \rm tomographic bins. Convolutions on {\sc healpix} maps are implemented as matrix multiplications on the {\sc healpix} arrays with the appropriate weights. This convolution matrix is precomputed once before training the neural networks.

Because our method uses the lognormal model as a starting point, we can use a physically motivated network architecture.  The same architecture (except the final layer) is shared by the generator and the discriminator. The network has 4 input channels, corresponding to the 4 tomographic input lognormal \healpix\ maps at $N_\mathrm{side}=512$ generated as described above.  The corresponding resolution is $\approx 7$~arcmin.  
We choose this scale based on the fact that on smaller scales baryonic physics become significant.  We use a ResNet \cite{resnet} architecture with 4 residual blocks and 8 channels per convolutional filter as detailed in Appendix \ref{app:nn_architecture}. 
The number of residual blocks is set by demanding that the receptive field size of the network match the causal horizon of each pixel ($\approx\ 0.5\ {\rm deg}$). As a result, the GAN only learns how to modify the lognormal map on small scales. The generator has four output channels, corresponding to each of the desired simulation-quality tomographic maps.  By contrast, the output of the discriminator is a single channel: i.e. the discriminator evaluates whether the set of 4 tomographic maps as a whole is distinguishable from simulations.

The details of the implementation of the rotationally equivariant convolutions and the residual block is presented in Appendix \ref{app:nn_architecture}.

\subsection{Network Training}\label{ssec:training}

Because the generator and discriminator only need to learn small-scale physics, it is unnecessary to train the networks on the full sky. Instead, we train the networks using small $3.5 \times 3.5\ {\rm deg}^2$ patches. To ensure the generator is not adversely affected by edge effects, the discriminator is restricted to the inner $2.5\times 2.5\ {\rm deg}^2$ patch, ensuring all pixels contributing to the discriminator include their full receptive field within the training patch. The number of pixels contributing to the discriminator is $\approx 400$ pixels per tomographic bin, or $\approx 1,600$ pixels in total.

To avoid recomputing the convolution matrix at each patch on the sky, we choose to train at a patch centered on the north pole. We include patches in arbitrary positions in the sky by rotating these patches onto the north pole at a resolution of $N_\mathrm{side} = 2048$. The patches are then downgraded to $N_\mathrm{side} = 512$. This procedure is performed at a higher resolution to avoid introducing resolution artifacts that become apparent when the rotation is done at $N_\mathrm{side} = 512$. As noted in section \ref{ssec:lognormal_model}, the same procedure is followed for the input lognormal maps. The spatial averaging introduces a pixel window function in both these maps. Therefore, our GAN learns to produce maps in which the value of the map at each pixel represents the pixel-averaged matter density.

The networks are trained by optimizing the hinge loss \citep{hinge_loss},
\begin{multline}
    L = \max_G \min_D \mathbb{E}_{\vec{\kappa}_{\text{sim}} \sim P_\mathrm{data}} [\min(0, D(\vec{\kappa}_{\text{sim}}) - 1)] \\- \mathbb{E}_{\vec{\kappa}_{\text{LN}} \sim P_\mathrm{prior}} [\min(0, -D(G(\vec{\kappa}_{\text{LN}})) - 1)].
\end{multline}
This loss is minimized relative to the discriminator parameters, but maximized relative to the generator parameters, where the optimal point in parameter space is a saddle point. The output of the discriminator is a number according to its assessment of the input as being real or fake. Because the generator's goal is to learn small perturbations to the lognormal model, we include an L1 identity loss term that limits the deviation of the generator mapping from the lognormal input, $\kappa_{\text{LN}}$,
\begin{equation}
     L_\mathrm{identity} = \mathbb{E}_{\kappa_{\text{LN}} \sim P_\mathrm{prior}} \left[| G(\vec{\kappa}_{\text{LN}}) - \vec{\kappa}_{\text{LN}}|_1 \right].
\end{equation}
The relative contribution of this identity loss term to the total loss is controlled by the hyperparameter, $\alpha$, so that the total loss is given by 
\begin{equation}
    L_{\rm tot}=L-\alpha L_{\rm identity}.
\end{equation}
The negative sign appears because the loss is maximized with respect to the generator parameters. Note that the additional terms in the loss function are only included for the generator training.    

Because GANs are difficult to train, much effort has gone into developing methods for improving their training stability \citep{wgan_gp, r1_gp,spec_norm}. We follow \cite{ttur} in training the discriminator with a higher learning rate than the generator but do not otherwise apply any other such regularization schemes. In particular, we have found that the small size of our networks combined with interpretable lognormal inputs and the $L_{\rm identity}$ penalty successfully stabilize the training.

\subsection{Synthetic map generation}\label{ssec:synthetic_map_gen}

As discussed in the previous section, we train the CNN on small patches at the north pole. Nevertheless, once trained, we can use the generator to produce full-sky synthetic mocks. 

As we discuss in Appendix \ref{app:nn_architecture}, our implementation of the CNN relies on finding the $4$ nearest {\sc healpix} pixels and then expressing the convolution as a matrix multiplication of the values from the neighboring pixels. 
An additional step that is involved in generating synthetic maps is to find the list of neighboring pixels that enters the convolution in equation \eqref{eqn:radial_conv}. We then use the same convolution weights derived by training on small patches to generate full-sky synthetic maps.  

As noted in Appendix \ref{app:nn_architecture} this procedure is only approximate since the distance between neighboring {\sc healpix} pixels can differ for different pixel pairs. Nevertheless, as we will see in the next section, the approximation is still able to learn non-Gaussian information in convergence maps.

\begin{figure*}
    \centering
    \includegraphics[width=\linewidth]{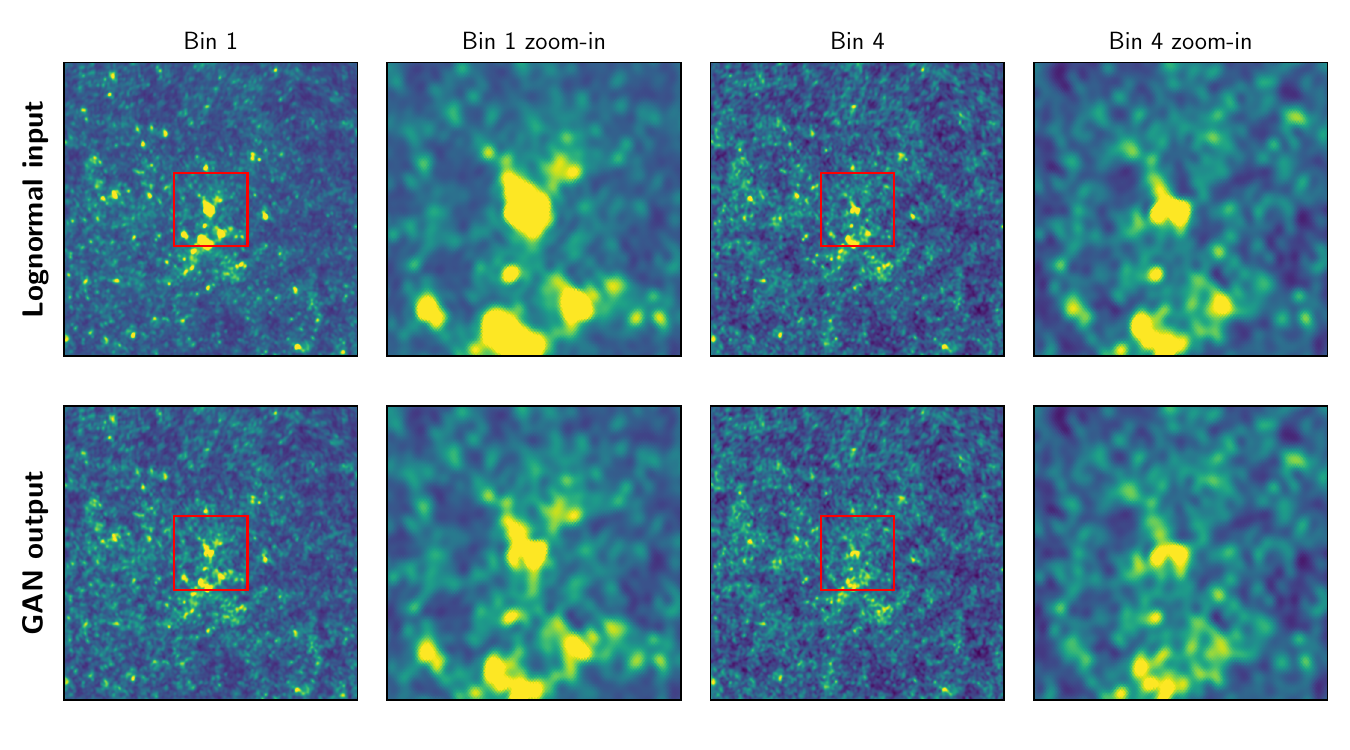}
    \caption{Comparison of a patch of the lognormal input (top row) and the output from our trained generator (bottom row) in the lowest (the two left columns) and highest (the two right columns) redshift bins centred on the most dense pixel of a \gansky map. The second and fourth panels are a zoom in of the red region indicated in the first and third panels. We see that the GAN outputs look very similar to the lognormal inputs. Looking at the zoom-in panels, we see that the density peaks of the lognormal input are made more compact by the GAN. Consequently, the GAN produces multiple peaks from a single peak in the lognormal input as seen in the density peak in the bottom of the zoom-in panels. The first and the third panels are ($15$ deg)$^2$ and the zoom-in panels are (3.5 deg)$^2$ in size. }
    \label{fig:map_plot}
\end{figure*}

\begin{figure}
    \centering
    \includegraphics[width=\linewidth]{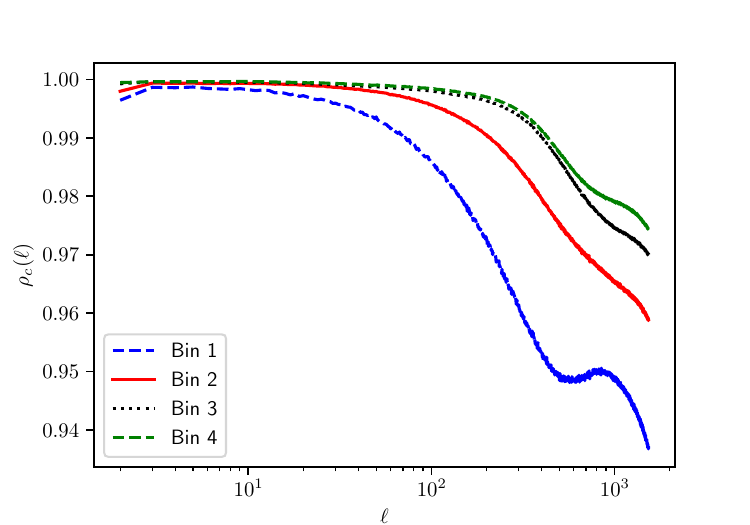}
    \caption{The cross-correlation between the \gansky output and its lognormal input maps. The different curves show the cross-correlation in different tomographic bins. As expected from the locality of the generator, the two maps are highly correlated. The cross-correlation at large scales (low $\ell$) approaches $1$, showing that the large scales are largely unaltered. Since the lognormal model describes the high redshift bins better than the low redshift bins, they have a higher cross-correlation than the low redshift bins. 
    }
    \label{fig:cross_corr}
\end{figure}

\begin{figure*}
    \centering
    \includegraphics[width=\linewidth]{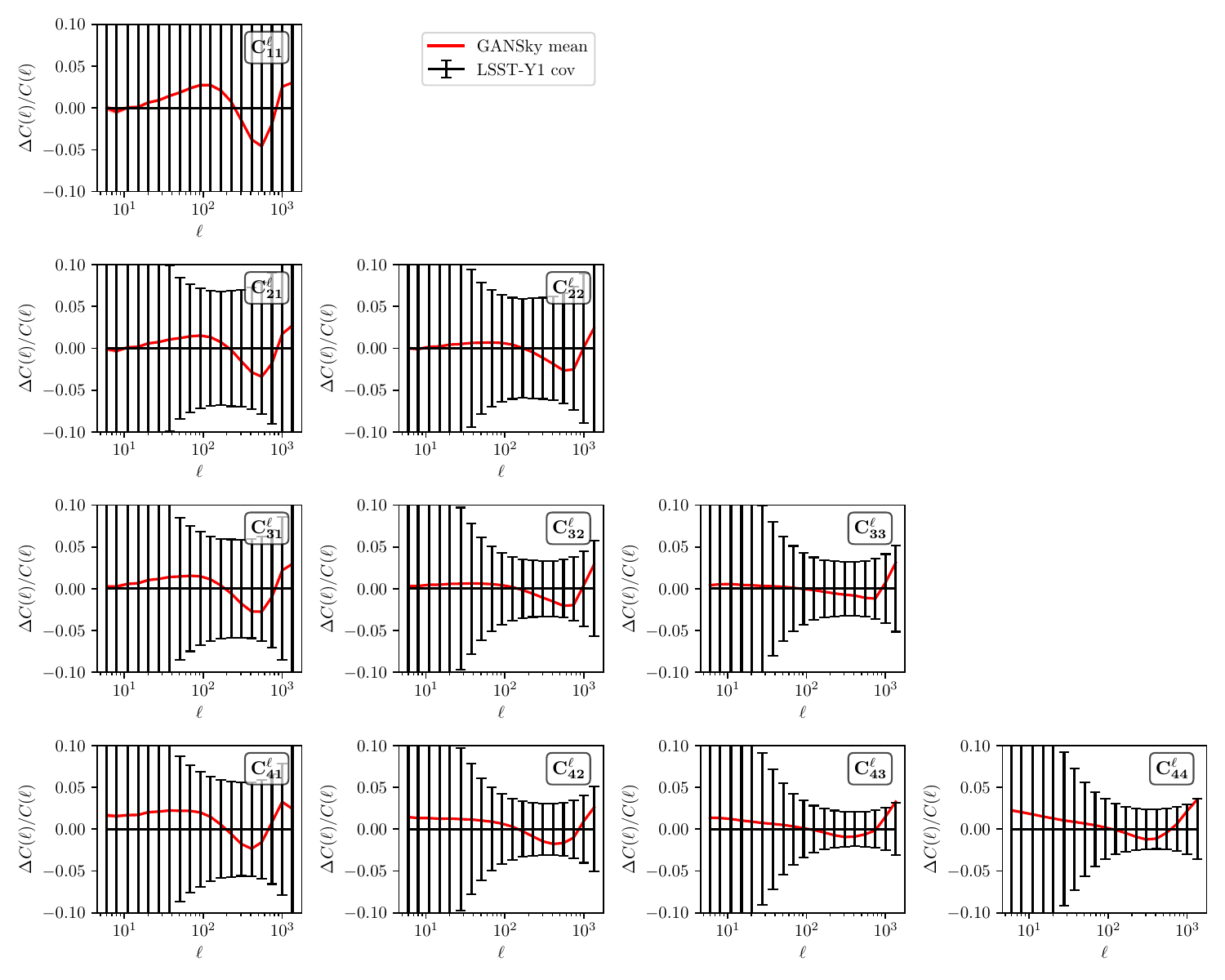}
    \caption{The relative difference in the power spectrum computed from the \gansky maps and the input lognormal maps. The different panels show the power spectrum between different tomographic bins as indicated in the plot. The red curve shows the fractional error in the mean of the power spectrum computed from $100$ \gansky mocks. The error bars show the uncertainty in the power spectrum for the LSST-Y1 survey scenarios. As we can see the error in the power spectrum is tolerable within the LSST-Y1 survey scenarios. See text for more details.
    }
    \label{fig:delta_Cl}
\end{figure*}

\begin{figure*}
    \centering
    \includegraphics[width=\linewidth]{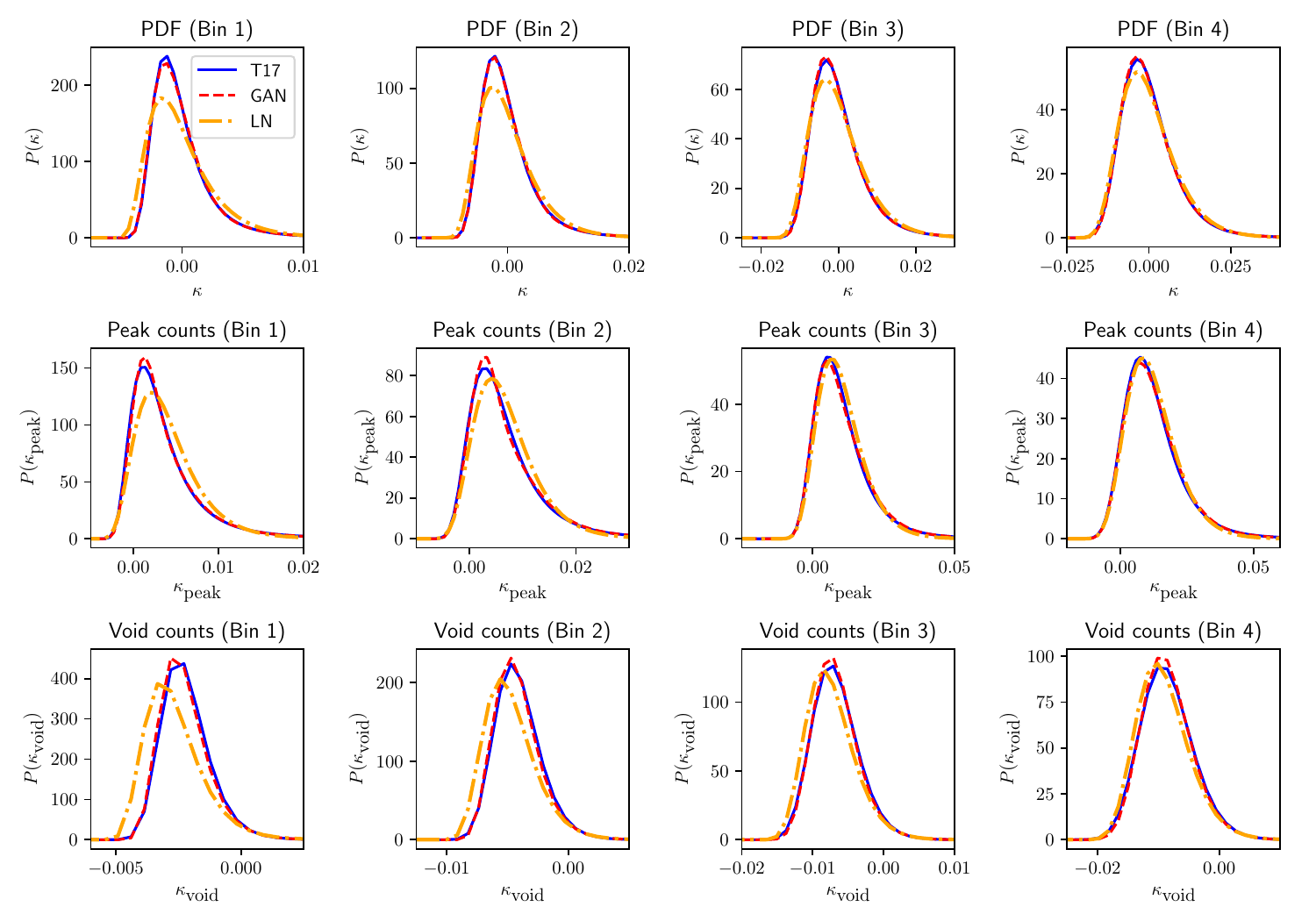}
    \caption{Comparison of non-Gaussian statistics computed from noiseless T17 maps (blue solid), \gansky maps (red dashed) and lognormal maps (orange dash-dotted). The three rows show three different non-Gaussian statistics -- PDF of $\kappa$ values (top row), peak counts (middle row) and void counts (bottom row) while the $4$ columns show the 4 different redshift bins. As we can see from the figure, the GAN maps reproduce all the non-Gaussian statistics of the ray-traced simulations. However the lognormal maps fail to reproduce the non-Gaussian summary statistics, especially at lower redshifts.}
    \label{fig:ng_stats}
\end{figure*}

\begin{figure*}
    \centering
    \includegraphics[width=\linewidth]{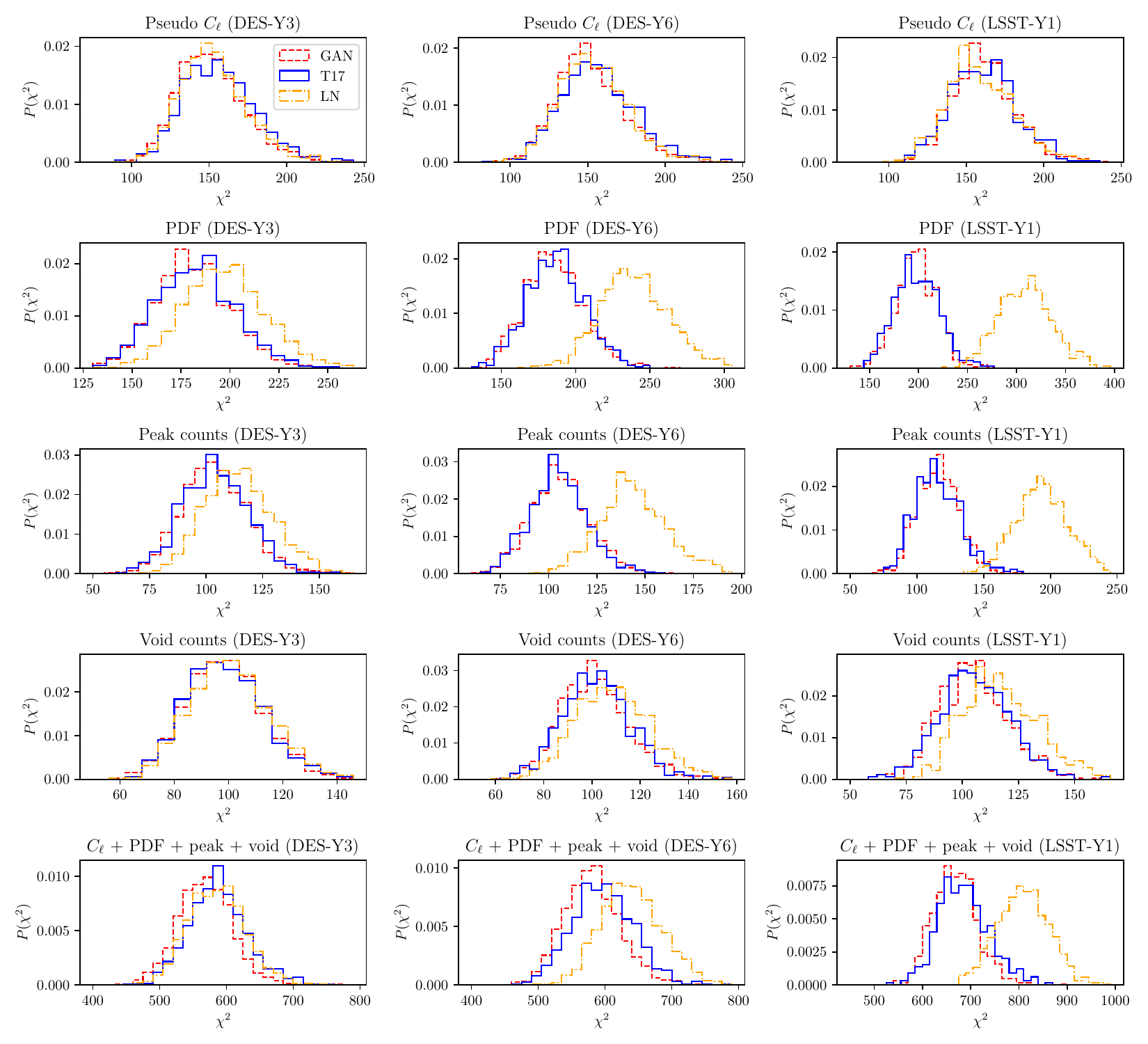}
    \caption{$\chi^2$ distribution of the different summary statistics -- pseudo $C_{\ell}$ (top row), $\kappa$ PDF (second row), peak counts (third row), void counts (fourth row), all the summary statistics combined (bottom row) for three different survey scenarios -- DES-Y3 (first column), DES-Y6 (middle column) and LSST-Y1 (right column). The $\chi^2$ for the mocks are computed relative to a mean and covariance computed from a set of \gansky mocks. The $\chi^2$ distribution of the ray-traced simulations are shown with solid blue histogram. GAN mocks are shown with dashed red histogram and the lognormal mocks are shown with dash-dot orange histogram. The GAN mocks are indistinguishable from the ray-traced simulations. On the other hand the $\chi^2$ distribution of non-Gaussian summary statistics from lognormal mocks are very different from that of the ray-traced simulation, especially for the DES-Y6 and LSST-Y1 survey scenarios. This shows that the GAN reproduces the second moment of the distribution of the non-Gaussian summary statistics and therefore we can use it for computing the covariance of these summary statistics.}
    \label{fig:chi_sq_dist}
\end{figure*}

\section{Results}\label{sec:results}

\subsection{Visual Inspection of the Maps}

Figure \ref{fig:map_plot} shows a patch of the lognormal convergence maps that are used as inputs (top row) to the corresponding \gansky output (bottom row) for the lowest and the highest redshift bin (Bin 1 and 4). The figure is centred on the densest pixel of a \gansky map. In the Figure we also show a zoom-in of the convergence peak. From the figure, we see the following: {\it i)} the GAN output looks remarkably similar to the lognormal input. Using interpretable lognormal convergence maps as the latent space means that the GAN only needs to learn how to perturb the input maps on small scales to mimic the ray-traced weak lensing simulations. {\it ii)} By focusing on the zoom-in regions, we can see that the action of the generator is to make the density peaks more compact. {\it iii)} Consequently, a single peak in the lognormal map can be broken into multiple peaks in the generator output. 

To quantify the similarity of the {\sc gansky} output with the lognormal inputs, we generated 100 pairs of lognormal and \gansky\ maps, and computed the correlation coefficient of the maps as a function of scale.  The latter is defined via
\begin{equation}
    \rho_c(\ell) = \frac{C_{\text{LN}\times \text{GAN}}(\ell)}{\sqrt{C_{\text{LN}}(\ell) C_{\text{GAN}}(\ell)}},
\end{equation}
where, $C_{\text{LN}\times \text{GAN}}(\ell)$ is the cross power spectrum between the input lognormal map and the output GAN map, while $C_{\text{LN}}$ and $C_{\text{GAN}}$ are the auto power spectrum of the lognormal and the GAN map respectively. The mean correlation coefficient as a function of scale for the different redshift bins is shown in Figure \ref{fig:cross_corr}. As we can see from the figure, the \gansky\ maps are tightly correlated with the input maps, with $\rho_c > 0.94$ at all scales and redshifts.  As expected, the correlation coefficient approaches unity on large scales.  We also note that the impact of the generator increases with decreasing redshift. This is because the impact of non-linear physics increases with decreasing redshift, so the input low redshift lognormal maps must be more heavily modified than their high redshift counterparts.

\subsection{Comparison of summary statistics}

To assess whether the accuracy of the GAN is sufficient for our purposes, we compare our generated sky maps to the T17 simulated sky maps for a variety of different summary statistics.

We first compare the power spectrum of the generated maps. While the input lognormal maps are designed to have the same power spectrum as the training simulations, there are no constraints in training to preserve the power spectrum under the action of the GAN. In Figure \ref{fig:delta_Cl}, we show the relative error between the auto and cross power spectrum in the $4$ redshift bins of the \gansky output and the input lognormal maps. The relative error in the power spectrum for all scales and across all redshift bins is $\lesssim 5\%$. In the figure, we also plot the uncertainty of the power spectrum assuming LSST-Y1 noise levels. The uncertainty shown in the figure is computed from the diagonal terms of the power spectrum covariance. We compute the power spectrum covariance analytically and only consider the `Gaussian' terms (equation A2 of \cite{Krause2017}) of the covariance. As we can see, the error in the power spectrum of the generated maps is lower than the uncertainty of the LSST-Y1 survey scenario.   

While the 2-point statistics of the lognormal input maps are the same as that of the T17 mocks by design, the same is not true for non-Gaussian statistics. Therefore, we now turn to investigate the non-Gaussian summary statistics. In Figure \ref{fig:ng_stats} we compare the 1 point PDF, peak counts and void counts of \gansky, T17 and lognormal maps. To compute the peak (void) counts, we identify local maxima (minima) of $\kappa$ values in pixels by comparing its values with all the neighboring pixels. The peak (void) counts are then defined as the distribution of $\kappa$ values of the peaks (voids). As we can see from the figure, the \gansky maps reproduce the above non-Gaussian summary statistics in the T17 simulations, whereas the lognormal maps fail to reproduce these statistics. The failure is most stark for the low redshift bins, where the impact of non-linear structure evolution is the strongest. Furthermore, while it is not apparent from the figure, the high end tails of the 1-pt PDF and peak counts cannot be reproduced by the lognormal maps.

\begin{table}
  \centering
  \caption{Survey configurations considered in this paper. $\sigma_{\epsilon}$ refers to the shape noise per ellipticity component. $n_{\text{eff}}$ is the effective number density in each tomographic bins and $f_{\text{sky}}$ is the fraction of sky in the survey footprint.}
  \begin{tabular}{l c c c}
  \hline
     Survey &  $\sigma_{\epsilon}$ & $n_{\text{eff}}$ [arcmin$^{-2}$] & $f_{\text{sky}}$\\
    \hline
    DES-Y3 & $0.26$ & $[1.476,1.479,1.484,1.461]$ & $0.125$\\
     DES-Y6 & $0.26$ & $[2.6175,2.6175,2.6175,2.6175]$ & $0.125$\\
     LSST-Y1 & $0.26$ & $[2.5,2.5,2.5,2.5]$ & $0.25$\\
    \hline
  \end{tabular}
  \label{tbl:survey_conf}
\end{table}

We wish to determine whether the differences between the simulated and \gansky\ maps are observationally significant for existing or near-future data sets. To that end, we compare a number of different summary statistics for noisy {\sc gansky}, lognormal and T17 maps at the noise levels of mock DES-Y3, DES-Y6 and LSST-Y1 surveys. The assumed properties of the different survey scenarios are given in Table \ref{tbl:survey_conf}. The DES-Y3 and DES-Y6 surveys are assumed to have a sky coverage of $5000$ deg$^2$ and the LSST-Y1 survey is assumed to be $10000$ deg$^2$ \cite{DESC_SRD}. For each of these survey scenarios, we use the same DES-Y3 redshift distribution shown in Figure \ref{fig:nz_plot}. In reality the redshift distribution for the DES-Y6 and the LSST-Y1 surveys will be different. For the purposes of our tests in this section, these scenarios should be thought of as simulated surveys with roughly the same statistical power of DES-Y6 and LSST-Y1. The effective number density for each survey scenarios is given in Table \ref{tbl:survey_conf}. The DES-Y3 number density is taken from \cite{DESY3_shapecatalogue}, and that of DES-Y6 and LSST-Y1 is taken from \cite{Fang2020}.

We compute the summary statistics considered in Figures \ref{fig:delta_Cl} and \ref{fig:ng_stats} from these noisy simulated mocks. We compute the power spectrum of the masked maps using the pseudo $C_{\ell}$ method \cite{Alonso2019}. We then calculate the $\chi^2$ associated with the difference between the two sets of mocks across the pseudo $C_{\ell}$, 1-pt PDF, peak counts, and void counts. We consider each statistic by itself, as well as all statistics simultaneously. To evaluate this $\chi^2$ we must calculate the full covariance matrix across the various summary statistics we considered.  Because we lack the large number of simulations required to do so, we rely on \gansky\ simulations to compute the covariance. Specifically, we generate $2000$ noisy \gansky\ maps at each of three noise levels we consider: DES-Y3, DES-Y6, and LSST-Y1. This is done by adding shape noise to the noiseless shear field of the \gansky\ mocks, and then transporting this noise back to the convergence field by using the standard Kaiser-Squires inversion \cite{Kaiser1993}. We then divide each full-sky mock into $4$ ($8$) patches for the LSST (DES) survey configurations corresponding to $f_{\text{sky}} = 0.25~(0.125)$. We compute the different summary statistics in each patch of the 2,000 simulations, and use these to compute the mean and covariance matrix for each of our summary statistics, including their joint covariance. For each patch, we then evaluate the corresponding $\chi^2$. We then repeat this procedure for a corresponding set of independently generated \gansky\ maps, as well as a corresponding set of lognormal maps.  We have explicitly verified that reducing the number of \gansky\ realizations used to calculate the covariance matrices has negligible impact on our results (specifically Figure~\ref{fig:chi_sq_dist}).

Figure~\ref{fig:chi_sq_dist} compares the $\chi^2$ distributions obtained using the T17 simulated maps and our \gansky\ maps. As we can see from the figure, the $\chi^2$ distribution from \gansky\ is indistinguishable for each of the survey assumptions.  Since the lognormal map have the same power spectrum by design, we do not see any difference in the $\chi^2$ distribution of pseudo $C_{\ell}$'s from the lognormal and T17 simulations. In contrast, we see clear difference between the $\chi^2$ distribution of lognormal maps and the T17 simulations for the non-Gaussian summary statistics, especially for the DES-Y6 and LSST-Y1 survey scenarios. The discrepancy for the lognormal mocks is highest for the PDF and peak counts, whereas the discrepancy of the $\chi^2$ distribution for the void counts is fairly small even in the LSST-Y1 scenario. From this comparison, we can see that \gansky\ maps reproduce not just the mean, but also the full $\chi^2$ distribution for each of our summary statistics. 

\begin{figure*}
    \centering
    \includegraphics[width=\linewidth]{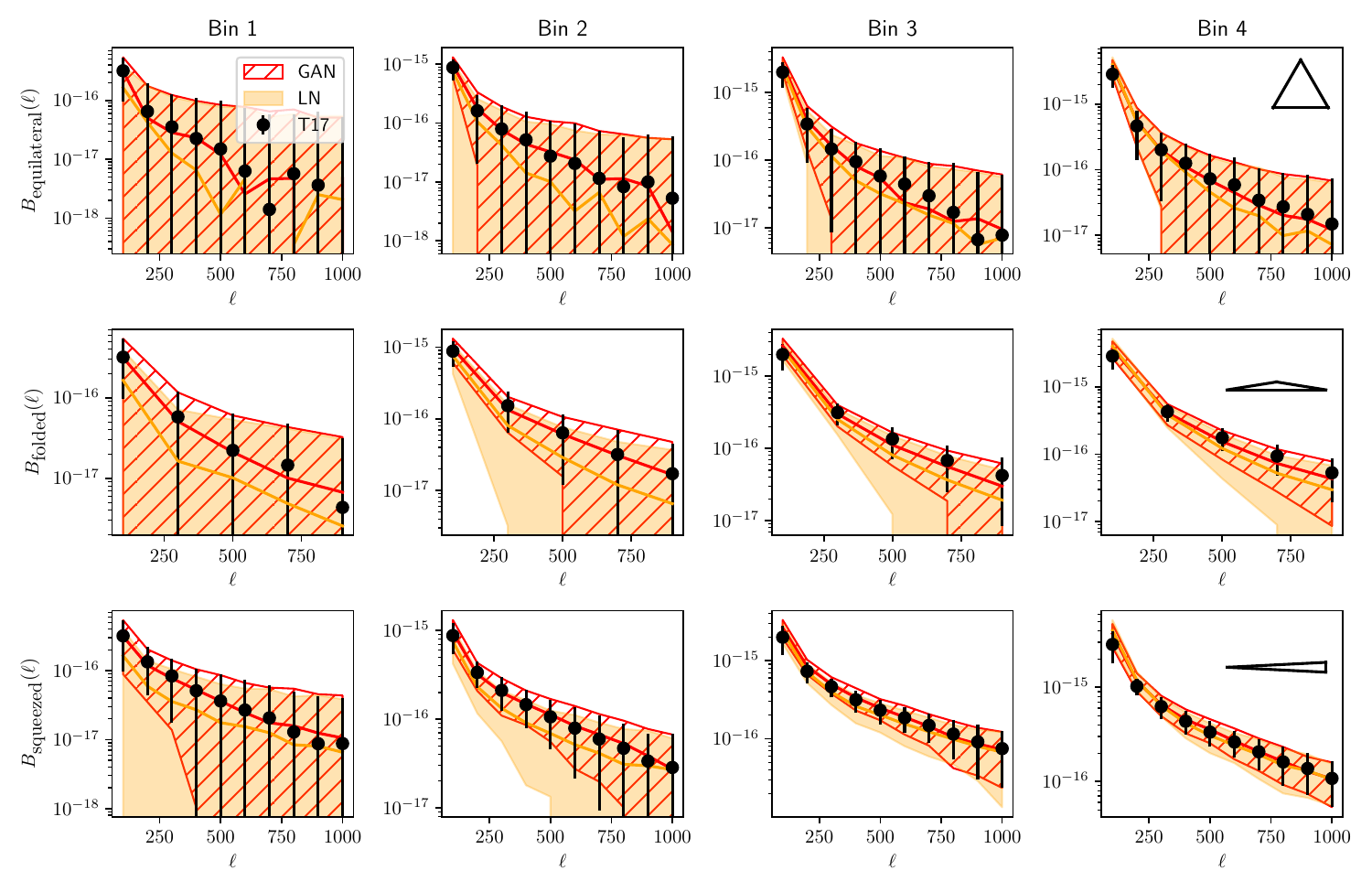}
    \caption{Bispectrum of T17 simulation maps (black error bars), lognormal maps (orange) and the \gansky maps (red hatch). The different columns show the bispectrum of the maps in different redshift bins. The top row shows the equilateral bispectrum, the middle row shows the folded bispectrum and the bottom row shows the squeezed bispectrum. The triangle configurations for each row is shown in the right column. The bispectrum noise is computed assuming LSST-Y1 noise levels. The GAN successfully reproduces the equilateral, folded and squeezed bispectrum in each redshift bins. }
    \label{fig:bispec_tomo}
\end{figure*}

\begin{figure*}
    \centering
    \includegraphics[width=\linewidth]{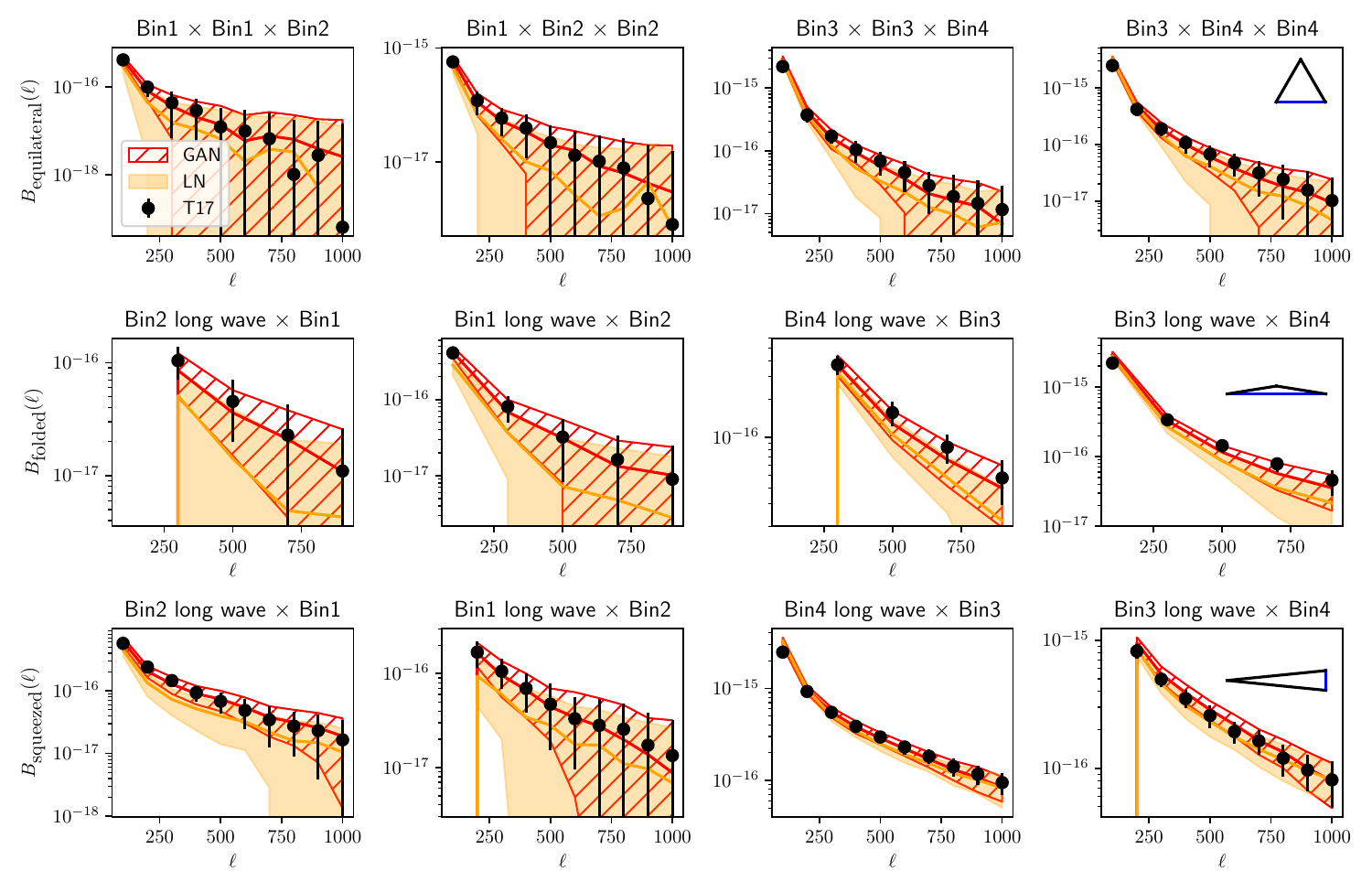}
    \caption{Similar to Figure \ref{fig:bispec_tomo}, but we consider cross redshift bin bispectra. The different redshift bin combinations are labelled in the title of each subplot. As we can see, we correctly recover the correct cross bispectra although the lognormal simulations do not have the correct cross bispectra.}
    \label{fig:bispec_cross}
\end{figure*}

\subsection{Comparison of Three-Point Statistics}

So far the non-Gaussian summary statistics we consider --- namely the 1-pt PDF, peak counts and void counts --- are `local' in the sense that they rely on pixel-by-pixel statistics. It is also useful to consider statistics that look at non-Gaussian correlations at various scales. Therefore, we compute the bispectrum, $B(\ell_1, \ell_2, \ell_3)$ of the lognormal, \gansky, and simulated maps using a binned bispectrum estimator \citep{Bucher_2010,Bucher_2016,Coulton_2019}. We compute the bispectrum in 10 linearly spaced $\ell$ bins between $\ell_{\text{min}}=50$ and $\ell_{\text{max}}=1050$ with a bin width of $\Delta\ell=100$ for $100$ noisy convergence maps with the DES-Y3/DES-Y6 and LSST-Y1 noise levels. Unfortunately, our estimator does not account for the survey mask, and is computationally expensive.  For this reason, we have chosen to compute the bispectrum mean and standard deviation using a set of 100 noisy full-sky maps produced using the procedure described in the previous section, except without a survey mask. We then scale the standard deviation by a factor of $1/f_{\text{sky}}$ to account for the different survey areas. In Figure \ref{fig:bispec_tomo}, we show three different configurations of bispectra for the LSST-Y1 noise level. In the top panels, we show the equilateral bispectrum, where $\ell_1 = \ell_2 = \ell_3$, for the $4$ redshift bins. Next, we show the folded bispectrum, where $\ell_1 = 2\ell_2 = 2\ell_3$ in the middle panels. Finally, in the bottom panels we show the squeezed bispectrum, where, $\ell_1 = \ell_{\text{min}}$ and $\ell_2 = \ell_3$. Note that in Figure \ref{fig:bispec_tomo} we only consider the auto bispectrum, where the three maps are from the same redshift bin. As we can see from the Figure, \gansky\ reproduces the equilateral, folded and squeezed bispectra on all scales and for all redshifts.

We also compute the bispectrum across different redshift bins. In weak lensing analyses, these cross-bin bispectra are effective at self-calibrating systematic parameters \cite[e.g,][]{Pyne2021}. Therefore, it is important to get the correct cross-bin non-Gaussian statistics. In Figure~\ref{fig:bispec_cross} we consider the correlation of one tomographic bin with its two adjacent redshift bins, i.e, Bin $i~\times$ Bin $i~\times$ Bin $(i\pm1)$.  The results are similar to those shown in Figure~\ref{fig:bispec_tomo}: \gansky successfully reproduces the equilateral, folded and squeezed bispectra.

\section{Conclusion}\label{sec:conclusion}

The science returns from Stage-IV surveys can be massively enhanced by extracting non-Gaussian information from these datasets. Data analysis to extract non-Gaussian information will require a large number of high-fidelity simulations. Given the limitations of existing methods (either accuracy or computational expense), we need new methods to produce weak lensing simulations that are simultaneously fast and accurate.

In this paper, we introduce an interpretable machine learning method to produce fast weak lensing simulations using Generative Adversarial Networks (GANs). Usually, the input to the generator of the GAN is a low-dimensional Gaussian/Uniform distribution. In our method, our latent variables are lognormal random fields with the same power spectrum and skewness as the maps that we try to emulate. Therefore, the generator only needs to alter the map on small scales to produce simulation quality maps. Due to the physically motivated input, we can get accurate maps with fairly small networks with $\mathcal{O}(10^3)$ parameters. Additionally, the action of our generator can be interpreted as a local mass-redistribution of the lognormal maps. We computed a number of summary statistics from the \gansky mocks and compared it to the same statistics from ray-traced simulations. We find that \gansky maps have the correct power spectrum, 1-pt function, peak counts and void counts. Not only does the GAN produce the correct mean for these summary statistics, but it also produces the correct second moment of their probability distribution.
Therefore, \gansky\ can be used for calculating the covariance for non-Gaussian summary statistics. Additionally, \gansky\ also reproduces the equilateral, folded and the squeezed bispectra of simulated maps. 

We anticipate these successes will enable us that can be used for the purposes of extracting non-Gaussian information from weak lensing data. However, there are some limitations and unexplored areas that will be addressed in the future. For example, in this paper \gansky\ was trained on simulations run at fixed cosmological parameters. For extracting cosmological information with non-Gaussianities we need to account for the cosmological dependence of the mapping between the lognormal and simulation-quality maps. Therefore, in the future, we will extend our method to produce weak lensing simulations with varying cosmological parameters. A straightforward way to extend our method to include cosmological parameters is to use a conditional GAN \citep[e.g, ][]{Perraudin2020}. An obstacle for training the GAN at varying cosmological parameters was the lack of full-sky weak lensing simulations for different cosmological parameters. Fortunately, over the past year several efforts for generating the necessary training data have been made public \citep[e.g.][]{Kacprzak2023, Gatti2023}. 

Another improvement that will be needed in the future is including the impact of baryons on the distribution of matter. Method similar to the one used here can be used learn the mapping from dark matter only simulations to maps including baryons. Indeed various analytical and machine learning methods have been proposed to map dark matter simulations to hydrodynamical simulations \citep{Dai2021, Horowitz2022, Boonkongkird2023, Sharma2024}. 

One of the main motivations for developing 
\gansky was to use it for field-level inference with weak lensing data. Indeed, machine learning emulators have been used within the field-level inference code {\sc borg} \cite{Jasche2013, Jasche2019} in \cite{Jamieson2023, Doeser2023}. We have previously developed the {\sc karmma} code that produces Bayesian mass maps using lognormal prior \cite{Fiedorowicz2023, Boruah2024}. 
In principle, we should be able to ``plug in'' \gansky\ into the \karmma\ framework to enable full-forward modeling of the observed shear field.
We will however need to test the performance of the GAN when sampling the high-dimensional mass maps with \gansky as part of the forward model. This will be investigated in a future work. 
\begin{table}
  \centering
  \caption{Different layers of the neural network architecture along with the number of input channels ($n_{\text{in}}$), the number of output channels ($n_{\text{out}}$) and the number of trainable parameters ($n_{\text{pars}}$). The generator ($G$) and the discriminator ($D$) have identical architecture except in the final layer.}
  \begin{tabular}{l | c | c | c}
  \hline
      Layer & $n_{\text{in}}$ & $n_{\text{out}}$ & $n_{\text{pars}}$\\
    \hline
     Initial $\mathcal{R}$ layer & $4$ & $8$ & $72$ \\
     Residual block 1 & $8$ & $8$ & $210$ \\
     Residual block 2 & $8$ & $8$ & $210$ \\
     Residual block 3 & $8$ & $8$ & $210$ \\
     Residual block 4 & $8$ & $8$ & $210$ \\
     Final $\mathcal{R}$ layer ($G$) & $8$ & $4$ & $68$ \\
     Final $\mathcal{R}$ layer ($D$) & $8$ & $1$ & $24$ \\
    \hline
    Total parameters ($G$) & & & $980$ \\
    Total parameters ($D$) & & & $936$ \\
    \hline
  \end{tabular}
  \label{tbl:nn_model_summary}
\end{table}

\begin{acknowledgments}
The authors are grateful to Shirley Ho for helpful discussions. The computation presented here was performed on the High Performance Computing (HPC) resources supported by the University of Arizona TRIF, UITS, and Research, Innovation, and Impact (RII) and maintained by the UArizona Research Technologies department. SSB was supported by the Department of Energy Cosmic Frontier program, grant DE-SC0020215, and NSF grant 200941. ER's work was supported is supported by NSF grant 2009401.  ER also received funding from DOE grant DE-SC0009913 and NSF grant 2206688. GF acknowledges the support of the European Research Council under the Marie Sk\l{}odowska Curie actions through the Individual Global Fellowship No.~892401 PiCOGAMBAS. The initial stages of this project were supported by the Center for Computational Astrophysics, Flatiron Institute.
\end{acknowledgments}

\appendix

\section{Details of neural network architecture and  implementation}\label{app:nn_architecture}

\begin{figure*}
    \centering
    \includegraphics[width=\linewidth]{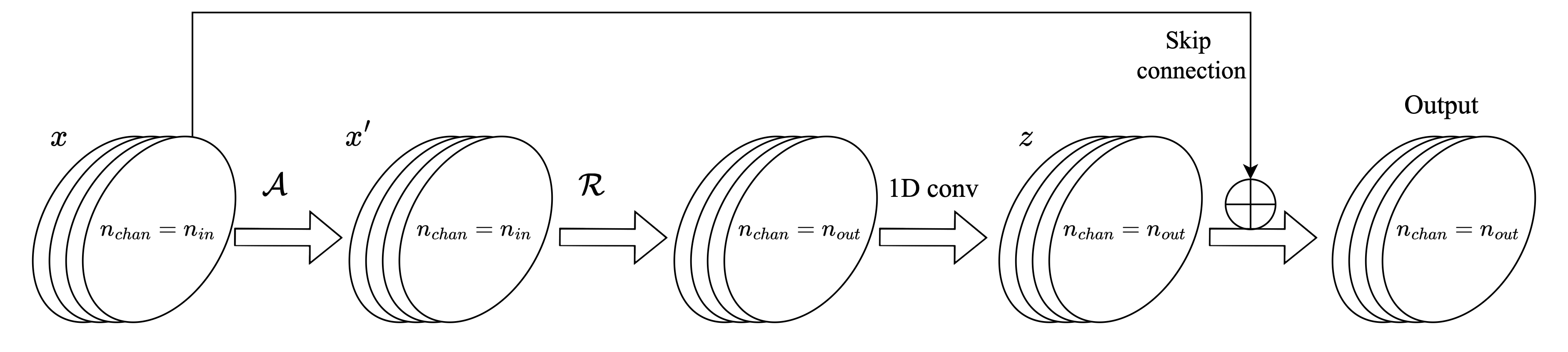}
    \caption{Illustration of the steps involved in each Residual block. Here $\mathcal{R}$ denotes one layer of the radial convolution and $\mathcal{A}$ denotes the application of the activation function. See text for more details.}
    \label{fig:GANSky_CNN}
\end{figure*}

In this Appendix we describe the details of our implementation of the neural network on {\sc healpix} maps. We implement rotationally equivariant radial convolutions on {\sc healpix} maps. It is implemented by first identifying the $4$ nearest neighbors\footnote{Pixels of a {\sc healpix} map contain 7 or 8 neighboring pixels with varying distances. We only use the $4$ nearest pixels for our convolutions.} of each pixel and then computing the linear combination of their values as, 
\begin{equation}\label{eqn:radial_conv}
    \mathcal{R}_i(x) = a x_i + b \sum_{j \in \mathcal{N}_i} x_j,    
\end{equation}
where, $\mathcal{N}_i$ denotes the set of the nearest neighboring pixels of the $i$-th pixel. The rotational equivariance is achieved by the equal contribution of the  neighboring pixels. The impositition of rotational equivariance in this work differs from that of   \cite{Perraudin2019}, who imposed rotational equivariance by pre-computing an adjacency matrix that depends on the distance between neighboring pixels. In contrast, our prescription for rotational equivariance is only approximate since the distance of the nearest {\sc healpix} neighbor depends on the location on the {\sc healpix} map. In our neural network, the initial and the final layers are a simple application of this radial convolutions. It is followed by the application of $4$ layers of a Residual block. The steps involved in each residual block is illustrated in Figure \ref{fig:GANSky_CNN}. The input of the residual block is first `activated' using a Leaky ReLU activation function followed by an application of $\mathcal{R}$ convolution. The same activation function is also applied to the output of the $\mathcal{R}$ convolution before an application of a cross-channel connection. In this connection, same pixels across different channels are connected to each other, but there is no cross-pixel. The output of the $k$-th channel of the cross-channel connection is given as, 
\begin{equation}
    z^{(k)} = \sum_{j=1}^{n_{\text{chan}}} w^{(k)}_{j} x_j + b^{(k)},
\end{equation}
where, $w^{(k)}$ and $b^{(k)}$ are the weights and biases associated with $k$-th output channel of the 1D convolution. Finally, a `skip' connection from the original input maps to the residual block is then added to the output from the cross-channel connection. We do not include any upsampling or downsampling layers. A summary of the different layers used in our neural network is given in Table \ref{tbl:nn_model_summary}.

\bibliography{gan_sky}

\end{document}